# A Study of the Occurrence of Supercooling of Water

Kah-Chye Tan [a)]

Wenxian Ho [b)]

*Addest Technovation, Singapore 139964*

J. I. Katz [c)]

*Department of Physics and McDonnell Center for the Space Sciences Washington University St. Louis, Mo. 63130 USA*

Shi-Jiang Feng [d)]

*Adducation, Guangzhou 510006, China*

## Abstract

Supercooling of water can be easily studied with a simple apparatus suitable for the student laboratory. We describe such an apparatus and its capabilities. The parameters influencing supercooling include the initial temperature of the water, and the temperature and the type of chilling medium. We correlate the occurrence of supercooling with the ability of the chilling medium to promptly nucleate ice; if it nucleates promptly, the layer of ice crystals formed on the boundary will initiate freezing of the bulk water without supercooling. If the chilling medium is unable to nucleate ice promptly, ice nucleation is delayed and the water supercools. Students can study and compare supercooling of distilled and natural water. Even quite dirty river water may be supercooled by as much as 5 °C.

# I. INTRODUCTION

Supercooling describes a substance that remains liquid when chilled below its freezing point. Pure water readily supercools well below its freezing point of 0 °C (273 K) at atmospheric pressure. Although supercooling was discovered[1] in 1724 and has been investigated extensively in the literature,[2-12] it is rarely discussed in science textbooks ranging from elementary to high school level. However, this phenomenon is of great practical importance because the freezing of supercooled water is responsible for the formation of slippery "glare ice" on ground surfaces, the breaking of ice-covered tree limbs, and the icing of aircraft surfaces. [13, 14]

Science teachers and students at the K-12 (pre-college) level should be aware of this remarkable phenomenon, which can be easily studied in the student laboratory. Another phenomenon, where hot water freezes before cold water (called the Mpemba effect), is closely related to the supercooling of water, and can also be studied at the K-12 and undergraduate levels.[11]

Our main interest here are the parameters that influence the occurrence of supercooling, such as the temperature of the water sample and the temperature of the medium chilling it.[9] Besides the temperatures of the water sample and the chilling medium, we also considered various types of chilling media, including frozen salt solution, refrigerated ethylene glycol solution, ice, and refrigerated air. By varying the type of chilling medium and its temperature, we explored a range of thermodynamic conditions.

In addition to studying the supercooling of distilled water, we also investigated whether impure water from a reservoir, a river, a sea, or rain can be supercooled.

## II. EXPERIMENTAL SETUP

Our experimental setup is called an IFM (Investigate Freezing and Melting) kit. It comprises an inner container attached to the middle of the base of an outer container, as shown in Fig. 1, and a plastic cover as shown in Fig. 2. The diameter of the inner container is 16.6 mm while that of the outer container is 65.0 mm. The volume of the outer container is 150 ml while that of the inner container is 10 ml. The water sample to be investigated is placed in the inner container while the liquid chilling medium is placed in the outer container.

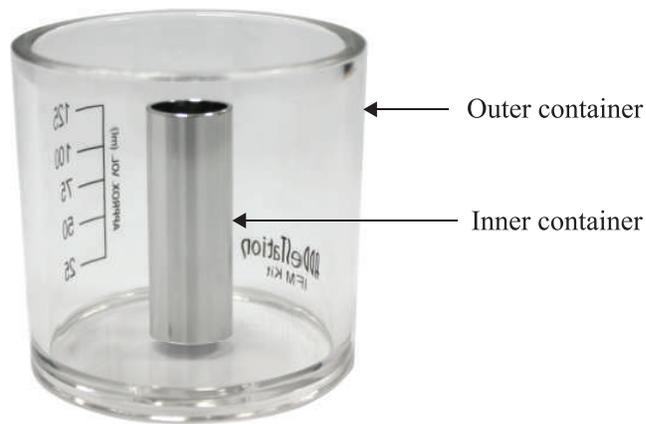

FIG. 1. Our experimental setup comprises an inner aluminum test tube and an outer plastic container. The water sample and the liquid chilling medium are placed into the inner container and the outer container, respectively.

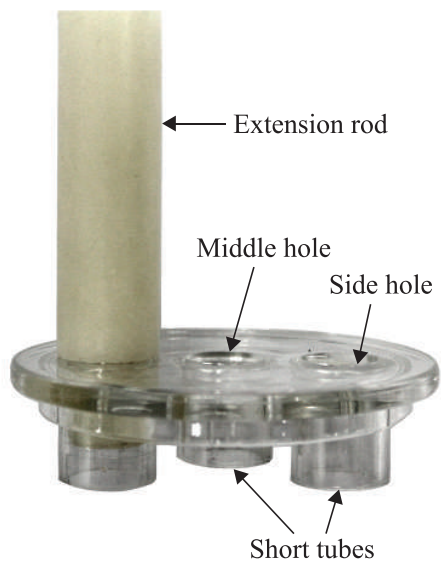

FIG. 2. The plastic cover has two holes for filling the containers, one in the middle and the other at the side. The short tubes below the holes prevent the water sample from mixing with the chilling medium. There is also an extension rod attached to the cover to facilitate the insertion of a temperature sensor.

The inner container is a thin (1.1 mm) aluminum test tube, a good conductor of heat. The outer container is made of thick (4.5 mm) plastic, a poor conductor of heat. Therefore, the rate of energy exchange between the water sample and the chilling medium through the metal tube is much greater than that between the chilling medium and the environment surrounding the outer container. Specifically, the characteristic thermal diffusion time $\tau = x^2 c\rho/\lambda$, where $x$, $c$, $\rho$, and $\lambda$ denote thickness, specific heat capacity, density, and thermal conductivity respectively, gives an indication of the rates of energy exchange through the metal tube and the outer container in relative terms. Using the physical properties listed in Table I, we find that $\tau = 188.6$ s for the plastic container while $\tau = 0.014$ s for the aluminum container.

TABLE I. Values of the thickness $x$, specific heat capacity $c$, density $\rho$, and thermal conductivity $\lambda$ of the plastic outer container and the metal tube.

| Material | Values | | | |
| --- | --- | --- | --- | --- |
|  | $x$ (m) | $c$ (J/kg °C) | $\rho$ (kg/m³) | $\lambda$ (W/m °C) |
| Plastic outer container | 0.0045 | 1500 | 1180 | 0.19 |
| Metal tube | 0.0011 | 897 | 2700 | 205 |

The plastic cover has two holes, one in the middle and the other off to the side, and fits over the outer plastic container. The metal tube is filled with the water sample through the middle hole and the outer container is filled with the liquid chilling medium through the side hole. Two short tubes prevent the water sample and the chilling medium from mixing. Meanwhile, a temperature sensor is inserted into the outer container through an extension rod on the cover to monitor the temperature of the chilling medium. This temperature sensor has a long plastic body and a metal tip that houses a thermistor; it therefore measures the temperature in a small region around the metal tip.

Once the liquid chilling medium has been placed into the outer container (the water sample is not to be placed in the metal container yet), the apparatus (together with the temperature sensor) is placed in a freezer. It is taken out of the freezer 24 hours later, when the chilling medium has cooled to the temperature of the freezer. The apparatus can then be used to freeze the water sample in the metal test tube without any external refrigeration.

In order to track temperature changes of the water sample during the trial, an extension rod is attached to the middle hole in the plastic cover and another temperature sensor is inserted into the metal test tube through the extension rod. The lengths of the extension rod and the temperature sensor have been determined so that the metal tip of the temperature sensor is positioned the same distance below the surface of the water sample in the metal test tube for every trial. It is essential to fix the depth at which the temperature sensor is submerged in order to minimize any undesired variations in the cooling curve, as discussed by Brownridge.[12] The two temperature sensors are then connected to an Addestation MGA datalogger[15] to obtain the temperature data.

### III.  CHILLING MEDIUM TEST

We carried out a preliminary study to determine the chilling medium to use. Three possible substances were identified: ethylene glycol solution, pure water, and a salt solution. Because the chilling medium is not refrigerated once it is taken out of the freezer, it is essential to find out how fast its temperature increases in a room-temperature environment. We want the temperature of the chilling medium to increase as slowly as possible during the course of each trial.

We placed 100 ml of ethylene glycol solution (50% concentration), pure water, and a salt solution (12% concentration) separately into the outer containers of three IFM kits that were then placed in a freezer maintained at –20 °C. (At –20 °C ethylene glycol solution is liquid, water is solid, and salt solution is a frozen slush of ice and liquid brine.) The three IFM kits were taken out of the freezer 24 hours later, and placed in a room-temperature environment. The experiment was repeated 10 times and we computed the 95% confidence intervals (based on the assumption of a normal distribution) for the temperature values at 360 s, 720 s, 1,080 s and 1,440 s. The temperature-time graphs of these three chilling media are shown in Fig. 3.

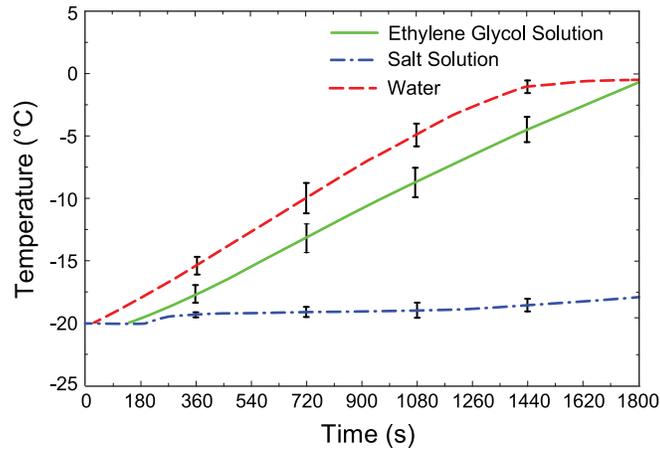

FIG. 3. Temperature-time graphs for 100 ml of a 50% ethylene glycol solution (solid), water/ice (dashed), and a 12% salt solution (dot-dashed). The solutions were placed in a freezer maintained at –20 °C for 24 hours and then moved to a room-temperature environment at $t = 0$. The experiment was repeated 10 times and the 95% confidence intervals (based on the assumption of a normal distribution) for the temperature values at 360 s, 720 s, 1,080 s and 1,440 s were computed and indicated on the graphs.

Because the data collection began when the temperatures of the cold media were consistently at –20 °C for all 10 trials, the temperature differences were expected to be small at the beginning. As time went by the temperature differences became larger, and thus the confidence intervals at 360 s are generally smaller than those at 720 s and 1,080 s. However, as the temperature increased further, the energy exchange between the cold medium and the surroundings became considerably less, causing the confidence interval to become smaller. This effect is particularly obvious for ice; even if its temperature in a trial increased much faster than the other trials, the temperature values started to level off as it approached 0 °C. The result is a smaller temperature difference towards the end of the experiment and hence a smaller confidence interval.

Based on these graphs it appears that the frozen salt solution is the best choice among the three types of chilling media considered as its temperature remains lower for much longer. However, because the chilling medium is a crucial component of the experiment, we conduct an additional investigation to confirm whether the frozen salt solution is indeed the best choice. In particular, we want to address two issues. First, why did the ethylene glycol solution and ice warm at comparable rates? Second, why did the frozen salt solution warm at a much lower rate than ice?

## IV. ACCOUNTING FOR THE WARMING TRENDS

### A. Why did the ethylene glycol solution and ice warm at comparable rates?

The ethylene glycol solution (50% concentration) remained a liquid throughout the experiment (its freezing point[16] is –36.8 °C). Due to convection, energy transfer to the temperature sensor will thus be more rapid in a liquid compared to a solid. However, the specific heat capacities of ethylene glycol solution (50% concentration) and ice are 3,266 J/(kg °C) [16] and 1,972 J/(kg °C), respectively. Thus, given the same energy flux through the outer container, the temperature of the ethylene glycol solution will rise more slowly compared to the ice. These two effects, a greater energy transfer efficiency and a greater heat capacity, counterbalance each other to some extent, thereby leading to a warming rate similar to that of ice.

### B. Why did the frozen salt solution warm at a much lower rate than the ice?

The temperature of the frozen salt solution changes from –20 °C to –16 °C in 30 minutes whereas it took only 3 minutes to change the temperature of the ice by the same amount. We believe this is due to the frozen salt solution having a much larger "effective" specific heat capacity, including the latent heat, than ice. However, we were not able to confirm this because applicable data on the heat capacity of frozen salt solution within the temperature range of interest was unavailable. We therefore carried out the following experiment to determine the ratio of the effective specific heat capacities of frozen salt solution and ice.

We first glued surface temperature sensors on the inner walls of two thin aluminum cups as shown in Fig. 4. Next, we filled one cup with salt solution and the other cup with water and covered each cup with a plastic lid in order to reduce energy transfer with the surroundings. We then placed the cups in a freezer maintained at –30 °C. The decrease in temperature was recorded using the datalogger for 10 independent trials.

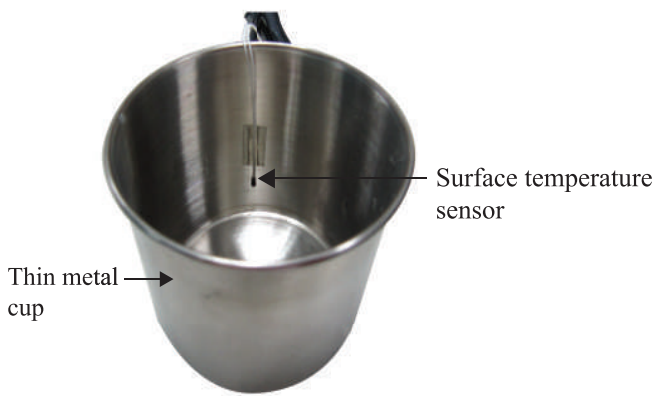

FIG. 4. A surface temperature sensor was glued onto the inner wall of a thin metal cup. Separate metal cups glued with surface temperature sensors were filled with salt solution and water, and plastic covers were placed on top of the cups. The setups were left in a freezer maintained at –30 °C.

Because the cups are thin and made of aluminium (a good conductor of heat), the temperature of the frozen salt solution (or water/ice) at the inner wall is practically the same as the temperature of the outer wall. Consequently, the temperature readings tracked the temperature of the frozen salt solution (or water/ice) in the metal cup as well as the outer surface temperature of the cup.

The surface temperature of the cup enabled us to estimate the amount of energy loss from the cup to its surroundings in the freezer. Based on Newton's Law of Cooling, the rate of energy loss of an object to the surroundings is

$$\frac{dQ}{dt} = k[T(t) - T_s], \qquad (1)$$

where $T(t)$ is the surface temperature of the object at time $t$, $T_s$ is the temperature of the surroundings (–30 °C in the freezer), and $k$ is a constant with units of W/°C. The value of $k$ depends on the surface area of the object being cooled, the physical properties of the fluid surrounding the object, and the nature of the convective flow.[17] In our experiment, the metal cup was cooled by the cold air in the freezer. The substance in the cup, be it frozen salt solution or water/ice, affected neither the surface area of the cup nor the physical properties of the cold air. In addition, the convective flow was not likely to vary much within the narrow applicable temperature range (–16 °C to –20 °C). Therefore, it is valid to use the same value of $k$ for the frozen salt solution and the ice in the experiment.

The accumulated heat loss $Q$ for the temperature to fall from –16 °C to –20 °C, the temperature range of concern, can be obtained from Eq. (1) as

$$Q = k \int_{t_1}^{t_2} [T(t) - T_s] \, dt, \qquad (2)$$

where $t_1$ and $t_2$ indicate the times at which the temperature of the outer wall is –16 °C and –20 °C, respectively, and $T_s = -30$ °C (the temperature of the freezer). The value of the integral can be obtained numerically from the measured cooling curve.

We use $Q=mc\Delta T$, where $m$ and $c$ denote the mass and the specific heat capacity of the substance under investigation, and $\Delta T = 4$ °C is the temperature change of concern. Substituting into Eq. (2), we find

$$c = \frac{k}{4m} \int_{t_1}^{t_2} [T(t) - T_s]\, dt. \tag{3}$$

The ratio of the effective specific heat capacity of frozen salt solution $c_F$ to that of ice $c_I$ is then

$$\frac{c_F}{c_I} = \frac{m_I}{m_F} \frac{\int_{t_1}^{t_2} [T_F(t) - T_s]\, dt}{\int_{t_1}^{t_2} [T_I(t) - T_s]\, dt}, \tag{4}$$

where $m_F$ and $m_I$ are the masses of frozen salt solution and ice, respectively, and $T_F(t)$ and $T_I(t)$ are their temperatures. Based on the 10 trials conducted, the average value of the integral for the frozen salt solution was 1097.6 °C min while that of ice was 256.9 °C min. The masses $m_F$ and $m_I$ were measured to be 128.8 g and 118.2 g respectively. The ratio of the effective specific heat capacities of frozen salt solution to that of ice was therefore 3.9.

The frozen salt solution has a large effective specific heat capacity because it contains regions with a range of salt concentrations that melt over a range of temperatures as it warms. The slushy nature of this mixture of solid and liquid can be readily observed. As the temperature of the frozen salt solution decreases, it releases a large amount of latent heat (of fusion) in addition to the sensible heat of the liquid and solid; the result is a large effective specific heat. In comparison, the decrease in the temperature of ice (always solid in this temperature range) involves only sensible heat and no latent heat.

The cooling curves of the salt solution and water obtained through the above experiment (see Fig. 5) support our hypothesis. Indeed, there was a "flat" interval in the cooling curve of water during which the water turned into ice. Shortly after this "flat" interval, the rate of decrease in temperature increased considerably. In contrast, there was no "flat" interval in the cooling curve of the salt solution and the temperature decreased at a slower rate from about –6 °C onwards. The salt solution did not turn into solid at a single temperature but froze gradually over more than 8 hours as it cooled along the ice/brine coexistence line. Consequently, latent heat and sensible heat were gradually released along the way. The salt solution did exhibit supercooling (which we will discuss in detail later) to a lower temperature as compared to the water, probably due to the presence of dissolved salt.

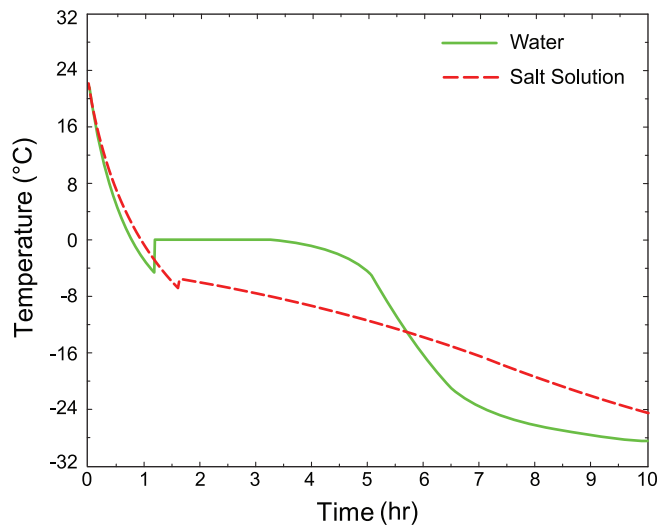

FIG. 5. The temperatures of the salt solution and water (kept in separate metal cups) were obtained using a surface temperature sensor glued to the inner wall of each of the cups.

On a separate note, we consistently observed a layer of frost on the surface of the IFM kit when frozen salt solution was used as the chilling medium (see Fig. 6). In comparison, frost was hardly ever seen when ethylene glycol solution or ice was used. This is to be expected as frozen salt solution has a considerably larger effective specific heat capacity and thus is able to absorb much more heat from the outer surface of the IFM kit, cooling it below 0 ºC. As a result, frost forms when the water vapor in air comes into contact with the cold surface. The presence of frost substantially reduces heat flow.[18]

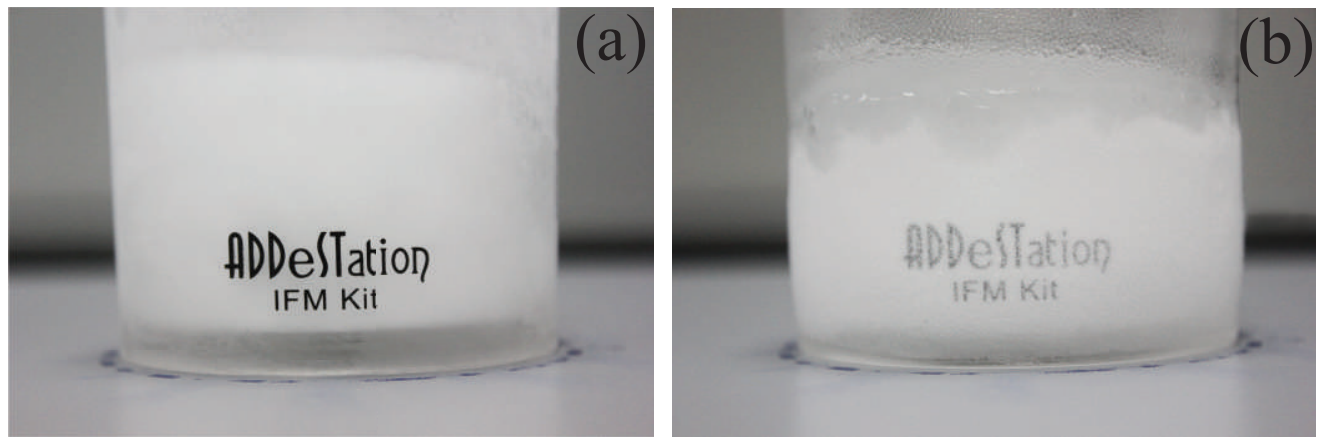

FIG. 6. (a) When the IFM kit containing frozen salt solution is taken out of the freezer, its outer surface is not covered with frost and the lettering can be clearly observed. (b) Shortly after removal, frost forms on the outer surface and the lettering becomes considerably less clear.

In summary, we understand why the temperature of the frozen salt solution rises much more slowly than the temperatures of ethylene glycol solution and ice. Because this is precisely the behavior we want, we will use the frozen salt solution as our choice of chilling medium.

## V. SUPERCOOLING DISTILLED WATER

In this section, distilled water was used for all experiments. The water was first boiled and then cooled to the desired temperature. The volume used for each trial was 6 ml. In addition to the frozen salt solution, we used refrigerated air as another chilling medium in order to reduce the cooling rate (discussed in Sec. V. C.).

### A. Frozen salt solution as chilling medium

We used 100 ml of frozen salt solution (12% concentration) as the chilling medium for all the experiments reported in this subsection. A total of 20 IFM kits with salt solution were put in a freezer maintained at –30 °C for at least 24 hours, allowing us to obtain many sets of data in a single day. Once each IFM kit was taken out of the freezer, we allowed the temperature of the frozen salt solution to increase to a specific temperature (–26 °C, –19 °C, –15 °C, –12 °C, or –8 °C) before working with it. We used three different temperatures for the water samples: 90 °C, 50 °C, and 23.7 °C (room temperature).

We conducted 10 trials for each pair of temperature values (chilling medium and water sample) and determined whether supercooling occurred for all trials. The criterion for supercooling is the presence of a dip in the cooling curve below 0 °C and a sudden rise in temperature, followed by a "flat" interval as shown in Fig. 7(a). The sudden rise in temperature is due to the release of latent heat associated with the onset of freezing. The water sample contains a mixture of ice and water during such a flat interval.

When water supercools, its temperature typically falls to around –5 °C. The temperature sensor used has an accuracy of 0.2 °C, and thus the measurement error will not affect our detection of supercooling. On the other hand, a typical cooling curve in the absence of supercooling is shown in Fig. 7(b). While there is still a flat interval, the cooling curve approaches this interval from above; it does not have the sudden rise shown in Fig. 7(a). The flat interval between approximately 180 s and 450 s in Fig. 7(b) is attributed to the phase transition as the water sample gradually freezes.

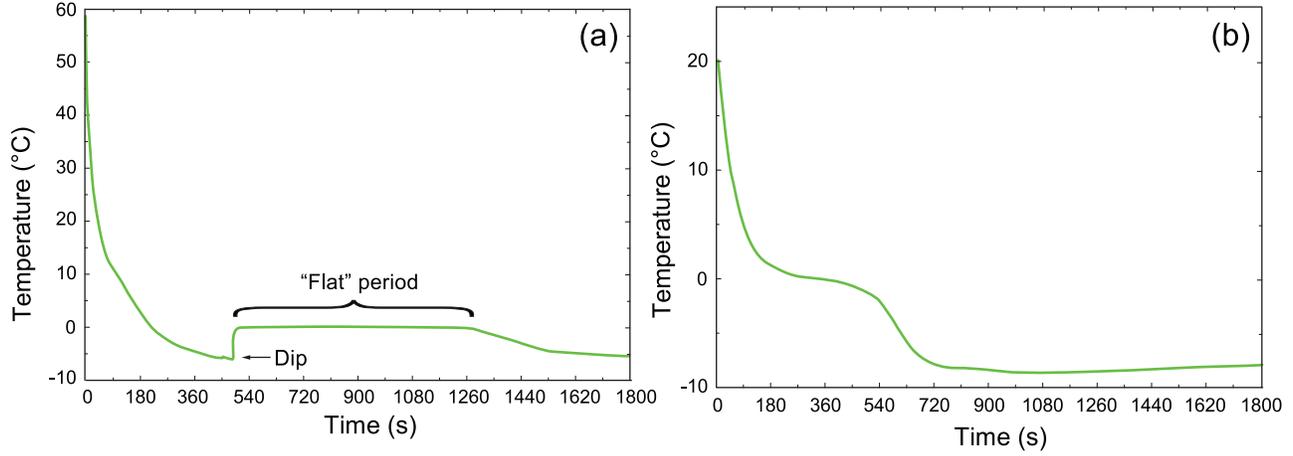

FIG. 7. (a) The cooling curve of a water sample that exhibited supercooling; it is characterized by a dip below 0 °C and then a sudden rise in temperature followed by a "flat" interval. (b) A typical cooling curve of a water sample that did not exhibit supercooling; it does not have a dip and a sudden rise as shown in (a).

We tabulate the number of times supercooling was observed in Table II. We note that the hotter (90 °C) water samples underwent significantly more supercooling than the cooler (23.7 °C) samples.

TABLE II. Number of times supercooling was observed out of 10 trials with frozen salt solution as the chilling medium. The water samples were distilled water.

| Temperature of frozen salt solution | Initial temperature of water sample | | |
| --- | --- | --- | --- |
| | 90 °C | 50 °C | 23.7 °C |
| −8 °C | 10 | 10 | 2 |
| −12 °C | 10 | 9 | 1 |
| −15 °C | 10 | 3 | 0 |
| −19 °C | 7 | 0 | 0 |
| −26 °C | 3 | 0 | 0 |

### B. Air as chilling medium

In this section we used refrigerated air at –18 °C as the chilling medium (the outer container of the apparatus is left empty) for all the experiments reported in this sub-section. The entire IFM kit was then put in a freezer maintained at –18 °C throughout each trial. As the metal tube was encased inside the outer (plastic) container, cold air in the freezer could not reach the metal test tube directly. Thus, energy exchange could only occur gradually through the air between the metal tube and the outer container.

Just as when the salt solution was used as the chilling medium, we used distilled water samples at 90 °C, 50 °C, and 23.7 °C. The IFM kits were left in the freezer throughout the trial and the temperatures of the water samples were monitored using the datalogger. A total of 10 trials were conducted and the number of times supercooling was observed is listed in Table III. In contrast to the previous results, we observed supercooling in all of the air-cooled samples.

TABLE III. Number of times supercooling was observed out of 10 trials with cold air as the chilling medium. The water samples were distilled water.

| Temperature of cold air | Initial temperature of water sample | | |
|---|---|---|---|
| | 90 °C | 50 °C | 23.7 °C |
| –18 °C | 10 | 10 | 10 |

### C. Combining the results presented in Tables II and III

The results presented in Table II involved the use of frozen salt solution at different temperatures as the chilling medium while those presented in Table III involved refrigerated air (at a single temperature) as the chilling medium. In order to find a useful single-parameter description of the effect of the chilling medium, we calculated the mean cooling rate as (Initial Temperature – 0 °C)/(Time Taken to Reach 0 °C). For each of the categories presented in Tables II and III, we computed the mean cooling rate for each of the 10 trials. We then averaged the mean cooling rates for the 10 trials and determined the 95% confidence intervals (based on the assumption of a normal distribution). These results are presented in Table IV.

TABLE IV. Average cooling rates in °C/min (based on 10 trials) for the various categories presented in Tables II and III and their 95% confidence intervals. The water samples were distilled water.

| Chilling medium | Initial temperature of water sample | | |
| --- | --- | --- | --- |
| | 90 °C | 50 °C | 23.7 °C |
| Cold air at –18 °C | 2.78 ± 0.30 | 1.72 ± 0.19 | 0.90 ± 0.12 |
| Frozen salt solution at –8 °C | 8.27 ± 0.81 | 7.93 ± 0.81 | 5.15 ± 1.53 |
| Frozen salt solution at –12 °C | 15.73 ± 1.39 | 13.67 ± 3.82 | 6.58 ± 1.49 |
| Frozen salt solution at –15 °C | 20.84 ± 2.31 | 15.38 ± 3.99 | 8.06 ± 0.75 |
| Frozen salt solution at –19 °C | 25.03 ± 6.29 | 19.84 ± 2.07 | 13.69 ± 1.60 |
| Frozen salt solution at –26 °C | 29.51 ± 8.02 | 21.52 ± 2.31 | 16.53 ± 1.76 |

The 95% confidence intervals show that the spread of cooling rates is reasonably small (around 10%) when all of the trials or none of the trials exhibit supercooling. When there is a mixture of supercooling and no supercooling, the spread of cooling rates is larger (around 25%). The larger spread in cooling rates is a result of the large difference in times taken for the temperature of the water sample to reach 0 °C when supercooling occurs compared to when no supercooling occurs.

We note that the cooling rate is much smaller when refrigerated air is used as the chilling medium compared to when the frozen salt solution is used as the chilling medium, even when the frozen salt solution is at a higher temperature. This behavior is expected due to the significantly higher heat capacity of the salt solution compared to refrigerated air. In addition, the cooling rate associated with the frozen salt solution is larger when its temperature is lower, in accordance with Newton's Law of Cooling. Likewise, when refrigerated air is used as the chilling medium, the cooling rate associated with a higher-temperature water sample is larger.

Another way of viewing the data presented in Tables II, III, and IV is to plot the number of times supercooling was observed against the mean cooling rate for each of the water samples (90 °C, 50 °C, and 23.7 °C). This plot is shown in Fig. 8.

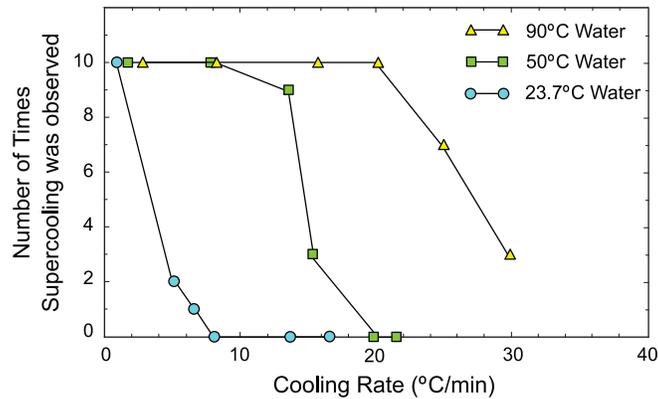

FIG. 8. The number of times supercooling was observed (see Tables II and III) was plotted against the mean cooling rate (Table IV) separately for the 90 °C, 50 °C, and 23.7 °C water samples.

## VI. ANALYSIS OF THE RESULTS

The results presented in Fig. 8 show that supercooling is highly probable when the water sample is initially very warm (90 °C) unless the cooling rate is very high. When the water sample is initially cooler (23.7 °C or 50 °C), supercooling only occurs when the cooling rate is correspondingly lower. That is, the lower the initial temperature of the water sample, the lower the cooling rate needs to be to observe supercooling.

To investigate how the temperature of the water sample and the cooling rate affected the occurrence of supercooling, we monitored the temperature of the water in contact with the metal tube during the entire cooling process. To do so, we glued a surface temperature sensor to the inner wall of the metal tube of the IFM kit. The sensor was positioned just above the water sample, so it is possible that the measurement would be affected by the temperature of the surrounding air. However, this effect should be small because the heat capacity of air is small compared to that of water.

Based on the results of Table II, we adopted a –15 °C frozen salt solution as the chilling medium and used a water sample initially at 90 °C to ensure supercooling, and also used a water sample initially at 23.7 °C to avoid supercooling. In addition to the temperature sensor glued to the metal tube, we also inserted a temperature sensor into the middle of the water sample immediately after the water was discharged into the tube.

Figure 9 shows a typical result for a water sample that is initially at 90 °C. The water sample cools very rapidly once it enters the metal tube, so by the time the temperature sensor is inserted and we begin taking data the temperature in the middle of the tube has already dropped to 37 °C. The data in Fig. 9 shows that supercooling occurs throughout the water sample. Moreover, the temperature readings at these two extreme locations both rose to their respective "flat" intervals at almost the same time, implying that the entire water sample remained supercooled until freezing began. We believe the slightly higher temperature of the sensor glued to the metal tube is caused by the surrounding air.

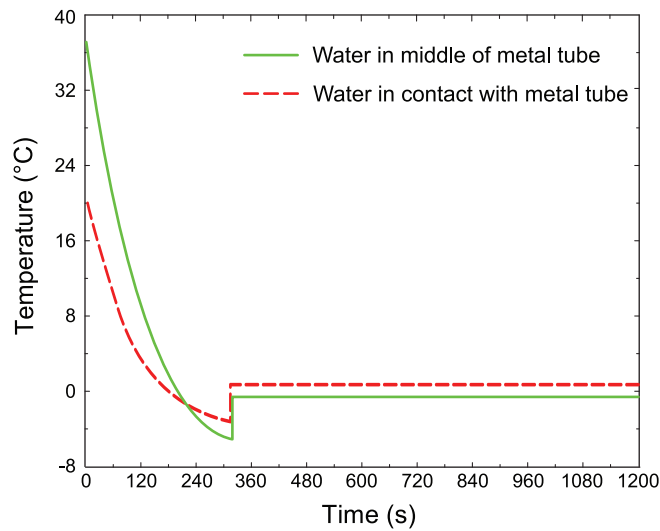

FIG. 9. The temperature readings of a water sample initially at 90 °C were obtained using a temperature sensor inserted into the middle of the metal tube and a temperature sensor glued to the inner wall of the metal tube. Rapid chilling by conduction through the thin metal tube cooled the middle of the water sample to 37 °C by the time the sensor was inserted.

Because the water was initially so warm, the temperature of the water in contact with the metal tube could only be reduced initially to about 20 °C under the influence of the chilling medium. Moreover, no ice was initially formed on the inner surface of the tube. We believe this lack of initial freezing is the reason this sample undergoes supercooling.

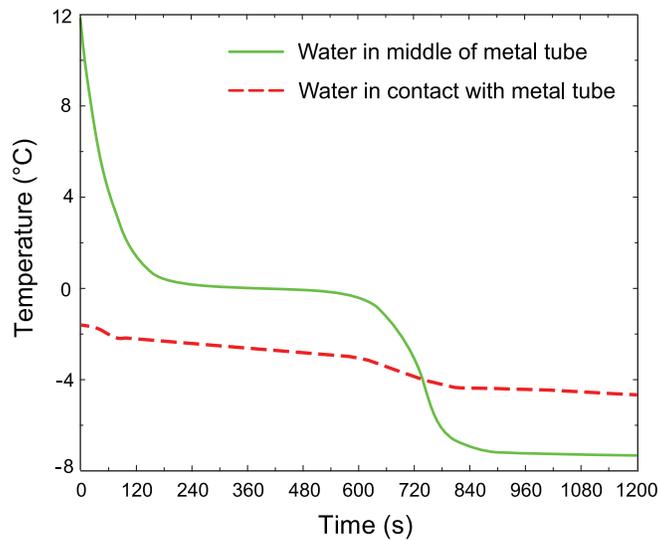

FIG. 10. The temperature readings of a water sample initially at 23.7 °C were obtained using a temperature sensor inserted into the middle of the metal tube and a temperature sensor glued to the inner wall of the metal tube.

Figure 10 shows a typical result for a water sample that is initially at 23.7 °C. The temperature of the metal tube was about –2 °C almost immediately after the water sample was placed into the metal tube. However, the temperature of the water in the middle of the tube was significantly higher, gradually decreasing to 0 °C at which point the typical "flat" interval associated with freezing appeared. Clearly, there is no supercooling in this instance, and freezing takes place starting from the boundary and expanding inwards. The formation of a layer of ice close to the inner wall of the metal tube with the middle of the water sample remaining liquid is evident in Fig. 11.

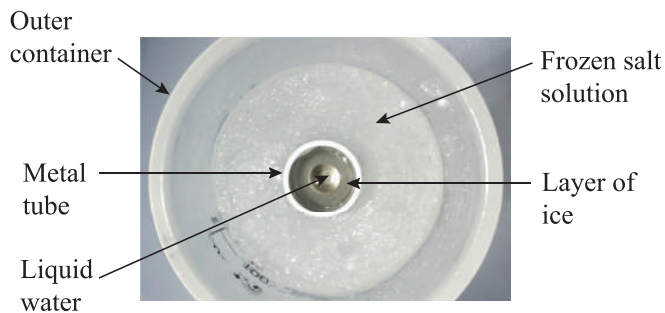

FIG. 11. Distilled water sample initially at 23.7 °C was discharged into the metal tube of an IFM kit with frozen salt solution at −15 °C as the chilling medium. A while later, a layer of ice close to the inner wall of the metal tube is observed while the water in the middle of the metal tube remains a liquid.

Our results suggest that supercooling is determined by whether the chilling medium is able to nucleate ice as soon as the water sample is introduced into the tube. If it does, the layer of ice crystals formed on the inner surface of the tube will initiate freezing of the bulk water without supercooling. Our experiments show that this happens when the water sample is comparatively cool or when the cooling rate is high. If the chilling medium is unable to nucleate ice immediately, no ice will form on the surface of the tube and the water has time to circulate and cool more uniformly. Without ice on the surface to initiate freezing, supercooling can occur. This happens when the water sample is warm or the cooling rate is low.

## VII. WILL NON-DISTILLED WATER SUPERCOOL?

In addition to distilled water samples, we also investigated water samples obtained from other sources such as a reservoir, a river, a sea, and rain. For both the river and sea water, we separately divided the water samples obtained into two groups. The first group was passed through a paper filter to get rid of the mud within, while the other group was left unfiltered. All in all, six types of water samples were tested: (a) reservoir water, (b) muddy river water, (c) filtered river water, (d) rain water, (e) muddy sea water, and (f) filtered sea water.

In each test, 6 ml of each water sample was preheated to 90 °C and discharged into the IFM kit using frozen salt solution at –15 °C as the chilling medium. As with the distilled water experiments, we performed 10 trials for each water sample. And just as with the distilled water experiment (see Table II), we observed supercooling in all 10 trials for every type of water sampled.

All the water samples supercooled by as much as 5 °C, including even muddy river and sea water that undoubtedly contained impurities that could provide freezing nuclei.[5] We therefore conclude that the suspended impurities in our samples do not preclude supercooling of water to –5 °C or so. Of course, it is not possible in these experiments to determine the nature of the impurities, which may even be biological,[19] that nucleate freezing.

## VIII. CONCLUSIONS

We considered the use of frozen salt solution, refrigerated ethylene glycol solution, ice, and refrigerated air as the chilling media for our experimental setup called the IFM kit. Frozen salt solution contains regions of varying salt concentrations that melt over a range of temperatures as it warms, giving it a much larger effective heat capacity as warming involves both latent heat as well as sensible heat. In comparison, warming ice (always solid) or ethylene glycol solution (always liquid) involve no latent heat.

Using frozen salt solution and refrigerated air, we subjected water samples at different temperatures to different rates of cooling, and tabulated the occurrences of supercooling under different conditions. Our experiments revealed that supercooling was most probable when the water sample was very warm unless the cooling rate was very high. When the water was cooler, supercooling rarely occurred unless the cooling rate was very low.

We went on to analyze the temperatures of the water at the boundary and in the middle of the metal tube with and without supercooling. Based on our analysis, we established a hypothesis that supercooling is determined by whether the chilling medium is able to nucleate ice immediately upon introduction of the water sample. If it does, the layer of ice crystals formed on the inner surface of the tube will initiate freezing of the bulk water without supercooling. If the chilling medium is unable to nucleate ice immediately, the water has time to circulate and cool uniformly, thereby permitting supercooling.

In addition to distilled water samples, we also studied the supercooling of water samples obtained from a reservoir, a river, a sea and rain. Although some of the water samples were quite muddy, all of them exhibited supercooling by as much as 5 °C. The supercooling of these water samples suggests that suspended impurities are not very effective in facilitating ice nucleation. This is unusual, as most metastable phase transitions are readily nucleated heterogeneously, and may be related to the large volume change of freezing water.

**ACKNOWLEDGMENTS**

We are grateful to Amir Gholaminejad of the Institute of Computational Engineering and Science, University of Texas at Austin, USA, for patiently answering all our questions regarding the experiments reported in the paper co-authored by him.[9]

______________________________________________

a) Electronic mail: kctan@addest.com

b) Electronic mail: wenxian@addest.com

c) Electronic mail: katz@wuphys.wustl.edu

d) Electronic mail: fengshijiang@sciedu.com.cn